\documentclass[10pt,letterpaper,twocolumn]{article}
\usepackage[english]{babel}
\usepackage{graphicx}

\newcommand{\alt}{\mathbin{\lower 3pt\hbox
   {$\rlap{\raise 5pt\hbox{$\char'074$}}\mathchar"7218$}}}
\newcommand{\agt}{\mathbin{\lower 3pt\hbox
   {$\rlap{\raise 5pt\hbox{$\char'076$}}\mathchar"7218$}}}

\begin{document}

\setcounter{footnote}{0}
\setcounter{equation}{0}
\setcounter{figure}{0}
\setcounter{table}{0}

\title{\large\bf Hidden Symmetry in 1D Localization}

\author{\small I. M. Suslov  \\
\small P.L.Kapitza Institute for Physical Problems,  \\
\small 119334 Moscow, Russia  \\
\small E-mail: suslov@kapitza.ras.ru\\
 {}\\
\parbox{120mm}{\footnotesize \,Resistance $\rho$
of an one-dimensional disordered
system of length $L$ has the log-normal
distribution in the limit of large $L$. Parameters of this
distribution depend on the Fermi level position, but are
independent on the boundary conditions. However, the boundary
conditions essentially affect the distribution of phases
entering the transfer matrix,
and generally change the  parameters of the evolution equation
for the distribution $P(\rho)$. This visible contradiction is
resolved by existence of the hidden symmetry, whose nature
is revealed by derivation of the equation for the stationary
phase distribution and by analysis of its transformation
properties.   } }

\date{}
\maketitle

\textwidth 6.4 in
\textheight 8.5 in

\setcounter{footnote}{0}
\setcounter{equation}{0}
\setcounter{figure}{0}
\setcounter{table}{0}

\begin{center}
{\bf 1. Introduction}
\end{center}

For description of 1D disordered systems
it is convenient to use the transfer matrix $T$, relating
the amplitudes of plane waves on the left ($Ae^{ikx}+Be^{-ikx}$)
and on the right ($Ce^{ikx}+De^{-ikx}$) of a scatterer,
$$
\left ( \begin{array}{cc} A \\ B \end{array} \right)\,
=  T \left ( \begin{array}{cc} C \\ D \end{array}\right)
\,.
$$
In the presence time-reversal invariance, the matrix $T$
can be parametrized in the form \cite{1}
$$
 T= \left ( \begin{array}{cc} \!\!\! 1/t\! &\! - r/t \!\!\\
\!\!- r^*/t^* \!&\! 1/t^* \!\!\!\end{array} \right)\,
= \left ( \begin{array}{cc}
\!\!\sqrt{\rho\!+\!1}\, e^{i\varphi}\!\! &
\!\!\sqrt{\rho} \,e^{i\theta}\!\!
\\ \!\!\sqrt{\rho}\, e^{-i\theta}\!\!
&\!\! \sqrt{\rho\!+\!1}\,
e^{-i\varphi}\!\! \end{array} \right)\,,
\eqno(1)
$$
where $t$ and $r$ are the amplitudes of transmission
and reflection, while $\rho=|r/t|^2$ is the dimensionless
Landauer resistance \cite{2}.
 For the successive
arrangement of scatterers their transfer matrices are
multiplied. For a weak scatterer its transfer matrix $T$ is
close to the unit one, which allows to derive the differential
evolution equations for its parameters, and in particular
for the Landauer resistance $\rho$.

In the random phase approximation (when distributions of
$\varphi$ and  $\theta$ are considered as uniform) such
equation for the distribution $P(\rho)$ has a form
\cite{3}--\cite{8}
$$
\frac{\partial P(\rho)}{\partial L} =
D\,\frac{\partial}{\partial \rho}
\left[\,\rho(1\!+\!\rho)\,\frac{\partial P(\rho)}{\partial \rho}
\,\right] \, \eqno(2)
$$
and describes evolution of the initial distribution
$P_0(\rho)=\delta(\rho)$ at zero length $L$ to the log-normal
distribution in the large $L$ limit.

As shown in the paper \cite{9}, the distributions of
phases $\varphi$ and $\theta$ change essentially, if
semi-transparent boundaries are introduced between the
disordered system and the ideal leads connected to it.
Then the more general equation arises
$$
\frac{\partial P(\rho)}{\partial L} =
D\,\frac{\partial}{\partial \rho}
\left[\,-\gamma(1\!+\!2\rho) P(\rho) +
\rho(1\!+\!\rho)\,\frac{\partial P(\rho)}{\partial \rho}
\,\right]   \,,
\eqno(3)
$$
which reduces to (2) in the random phase approximation.
The latter approximation is working sufficiently good in
the deep of the allowed band for the "natural" ideal
leads (made from the same material as a disordered
system, but without impurities),
as it is usually accepted in the theoretical papers (see
references in \cite{10,10a,11}); a situation in the forbidden
band is considered infrequently \cite{12,13,14} and only on the
level of the wave functions.  To study the evolution
of $P(\rho)$
for the arbitrary Fermi level position  (including the
forbidden band), one should explicitly introduce the foreign
ideal leads made from the good metal; as a result, the still more
general equation arises \cite{15},
$$
\frac{\partial P(\rho)}{\partial L} =
D\,\frac{\partial}{\partial \rho}
\left[\,-\gamma_1(1\!+\!2\rho) P(\rho)-
\vphantom{\frac{1}{2}} \right.
$$
$$  \left.
\,-2\gamma_2 \sqrt{\rho(1\!+\!\rho)} P(\rho)
+\rho(1\!+\!\rho)\,\frac{\partial P(\rho)}{\partial \rho}
\,\right]   \,,
\eqno(4)
$$
whose coefficients are determined by the stationary phase
distribution (see Eqs.\,29,\,31 below) in the large $L$ limit.
Equation (4) reduces to (3) with $\gamma=\gamma_1+\gamma_2$
in the region of large $L$, when typical values of $\rho$ are
large. Meanwhile, it
becomes clear that the random phase approximation is violated due
to internal reasons and a change of the boundary conditions is
not essential
for it. According to the paper \cite{15}, the distribution
$P(\rho)$ in the limit of large $L$ has the log-normal form
$$
P(\rho)=\frac{1}{\rho \sqrt{4\pi D L}}
\exp\left\{-\frac{[\ln \rho-vL]^2}{4DL}\right\}\,
\eqno(5)
$$
with $v=(2\gamma\!+\! 1)D$, whose parameters are determined by
the internal properties of the system, and does not depend on the
boundary conditions.  Fig.1 illustrates the dependence of the
parameter $\gamma$ on the quantity  ${\cal E}/W^{4/3}$, where
${\cal E}$ is the Fermi energy counted from the lower edge of the
initial band, and $W$ is the amplitude of a random potetial;
all energies are measured in the units of the hopping integral
for the 1D Anderson model (see below Eq.9). One can see that the
parameter $\gamma$ is formally always finite but takes small
values in the deep of the allowed band, in correspondence with
the random phase approximation.

\begin{figure}
\centerline{\includegraphics[width=3.2 in]{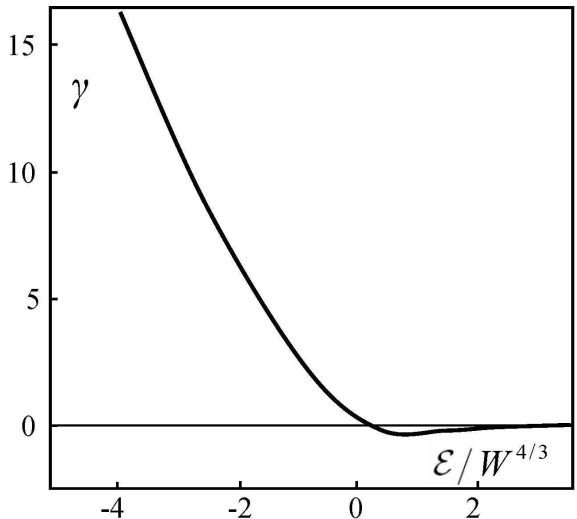}}
\caption{ Parameter $\gamma$ in equation (2),
corresponding to the limit of large $l$, as function
 of the energy ${\cal E}$,
counted from the lower edge of the initial band.
} \label{fig1}
\end{figure}
%

One can see that two statements were made in the papers
\cite{9,15}, which look hardly compatible. On one hand,
variation of the boundary conditions essentially affects
the distribution of phases, which generally
changes the parameters
of the evolution equations (2--4) and even its structure. On
the other hand, these changes have no influence on the form of
the limiting distribution (5) in the large $L$ region.
Validity of these two statements means that the system obeys
a hidden symmetry, i.e. invariance of the physical quantities
respective to a certain class of transformations. From the
theoretical viewpoint, revelation of the hidden symmetry
is of the evident interest, indicating the possibility of
essential simplifications. From the practical point,
one cannot differ the real physical effects from the fictive
ones, if the nature of hidden invariance is not clarified.
Revelation of this invariance appears to be very nontrivial
and we demonstrate it below for a set of
transformations discussed in Ref. \cite{15} and related
with a change of properties of the ideal leads attached to the
system.

Let explain the origin of two indicated statements. Under
a change of the boundary conditions, the transfer matrix $T$
transforms to $\tilde T=T_l T T_r$, where $T_l$ and $T_r$
are the edge matrices, related amplitudes of waves on the left
and on the right of the corresponding interface. Thereby, the
change of the boundary conditions leads to the linear
transformation of the trasfer matrix elements. The linear
transformation does not affect the growth exponents
for the second and forth moments of the matrix
elements, which can be found for a given matrix $T$ and
hence are
determined by internal properties of the system. Knowledge
of these two exponents allows to establish "the diffusion
constant" $D$ and  "the drift velocity" $v$ in the limiting
distribution (5), which consequently does not depend on the
boundary conditions \cite{15}; in particular, the behavior of
the parameter $\gamma$ (Fig.1) was established in such
way.\,\footnote{\,In this approach, the problem of phase
distribution was completely avoided. Of course, the same results
can be obtained by solution of Eq.31 and calculation of
averages in Eq.29. }

Influence of boundary conditions on the  distribution of phases
can be easily demonstrated by introducing the point scatterers on
the system boundaries, when
$$
\tilde T=T_l T T_r\,,\qquad
T_l=T_r=
\left ( \begin{array}{cc} 1\!-\!i\chi & -i\chi \\
         i\chi & 1\!+\!i\chi \end{array} \right)\,.
\eqno(6)
$$
Accepting the parametrization (1) for $\tilde T$,
one has in the main order for large $\chi$
$$
\sqrt{1\!+\!\rho} \,{\rm e}^{i\varphi} = -\chi^2\,  T'\,, \qquad
\sqrt{\rho}\, {\rm e}^{i\theta} = -\chi^2\,  T'\,,
$$
$$
\sqrt{\rho}\, {\rm e}^{-i\theta} = \chi^2 \, T'\,, \qquad
\sqrt{1\!+\!\rho}\, {\rm e}^{-i\varphi} = \chi^2 \, T'\,,
\eqno(7)
$$
where $T'=T_{11}\!-\!T_{12}\!+\!T_{21}\!-\!T_{22}$ and $T_{ij}$
are the elements of the $T$-matrix.
For large $\chi$ we have $\rho\sim \chi^4$ and
$1\!+\!\rho \approx \rho$, so it is easy to see
that
$$
\varphi =\pm\pi/2, \quad \theta =\pm\pi/2  \qquad
\mbox{\rm for}\,\quad \chi \to \infty\,.
\eqno(8)
$$
Thereby, for large $\chi$  the phase variables $\varphi$ and
$\theta$ are localized near values $\pm \pi/2$ independently
of their distributions in the initial system.

Let discuss the character of invariance mentioned above.
The change of the matrix $T$ with a system length $L$
is determined by relation  $T_{L+\Delta
L}=T_{L}\,T_{\Delta L}$, where the matrix $T_{\Delta
L}$ is close to the unit one; it allows to derive the
differential evolution equations. For the change of boundary
conditions, let multiply this relation by $T_l$ and $T_r$,
introducing the product $T_r T_r^{-1}=1$ between two multipliers;
then the analogous relation
$\tilde T_{L+\Delta L} = \tilde T_{L} \, T'_{\Delta L}$
arises for the matrix $\tilde T_{L}=T_{l} T_ L T_{r}$, where
the matrix $T'_{\Delta L}=T^{-1}_{r}T_{\Delta L}T_{r}$
is again close to the unit one. A passage from
$T_{\Delta L}$ to $T'_{\Delta L}$ changes the form of the
evolution equations, while a passage from  $T_{L}$ to $\tilde
T_{L}$ changes the stationary phase distribution, which
determines coefficients in Eq.\,4 for $P(\rho)$.  These two
factors should compensate each other, in order the limiting
distribution $P(\rho)$ remains invariant. However, such
invariance is not evident from equations, and
the general analysis of the situation looks
problematically at the present time. Below we restrict
ourselves by the partial case, when variation of the
boundary conditions is related with the difference of the Fermi
momentum $k$ in the ideal leads from the Fermi momentum
$\bar k$ in the system under consideration \cite{15}.


\begin{center}
{\bf 2. Initial relations}
\end{center}

As clear from experience of the paper \cite{15}, it is
convenient to consider the energies incide the forbidden
band of the initial crystal, while the description of the
allowed band can be obtained by analytical continuation.
For definiteness, we have in mind the 1D Anderson model
$$
\Psi_{n+1}+\Psi_{n-1}+V_n \Psi_n = E \Psi_n \,
\eqno(9)
$$
near the band edge, where it corresponds to discretization
of the usual continuous Schroedinger equation.

A scatterer in the forbidden band is described by the
pseudo-transfer matrix $t$, relating solutions on the left
($Ae^{\kappa x}+Be^{-\kappa x}$) and on the right
($Ce^{\kappa x}+De^{-\kappa x}$) of the scatterer.
Succession of scatterers with amplitudes
$V_0$,\,$V_1$,\,$V_2$,\,$\ldots$,\,$V_n$, arranged at
the points $0$, $L_1$, $L_1\!+\!L_2$,
$\ldots\,\,$, $L_1\!+\!L_2\!+\!\ldots\!+\!L_n$, is described
by the matrix
$$
t^{(n)}= t_{\epsilon_0} \, t_{\delta_1} \, t_{\epsilon_1}\,
t_{\delta_2}\,t_{\epsilon_2}\,
\ldots\,
 t_{\delta_n}\, t_{\epsilon_n}\,,
\eqno(10)
$$
where
$$
 t_{\epsilon_n}=
\left ( \begin{array}{cc} 1+\bar\epsilon_n & \bar\epsilon_n \\
-\bar\epsilon_n & 1-\bar\epsilon_n \end{array} \right)\,,\qquad
\bar\epsilon_n =\frac{V_n}{2\kappa a_0}\,,
\eqno(11)
$$
$$
t_{\delta_n}= \left ( \begin{array}{cc}
{\rm e}^{-\delta_n} & 0 \\ 0 & {\rm e}^{\delta_n} \end{array}
\right)\,,\qquad \delta_n =\kappa L_n\,
$$
and $a_0$ is the lattice constant. To obtain the true transfer
matrix $T^{(n)}=T_l\,\, t^{(n)} \,T_r$, we use the edge matrices
$$
T_l =  \left ( \begin{array}{cc} l\,\, & l^* \\
l^* & l\,\, \end{array} \right)\,,\qquad
T_r= \left ( \begin{array}{cc} r\,\, & r^* \\
r^* & r\,\, \end{array} \right)\,,
\eqno(12)
$$
$$
l=\frac{1}{2}\left(1+\frac{\kappa}{ik} \right)\,,\qquad
r=\frac{1}{2}\left(1+\frac{ik}{\kappa} \right)\,,
$$
describing the attachment of the
ideal leads made from the good metal with the Fermi
momentum $k$. Introducing the
product $ T_r T_l=1$ between any two multipliers in Eq.10, we
have
$$
 T^{(n)}= T_{\epsilon_0}
\, T_{\delta_1} \, T_{\epsilon_1}\,
T_{\delta_2}\,T_{\epsilon_2}\,
\ldots\,
 T_{\delta_n}\, T_{\epsilon_n}\,,
\eqno(13)
$$
where
$$
T_{\epsilon_n}=  T_l \,\,t_{\epsilon_n}\,  T_r \,,\qquad
T_{\delta_n}=  T_l \,\,t_{\delta_n}\,  T_r\,.
\eqno(14)
$$
In the Anderson model all $\delta_n$ are equil,
$\delta_n=\kappa a_0$, since the scatterers are present at each
site of the lattice. Substituting (12), we have
$$
 T_{\epsilon_n}=
\left ( \begin{array}{cc} 1\!-\!i\epsilon_n & -i\epsilon_n \\
i\epsilon_n & 1\!+\!i\epsilon_n \end{array} \right)\,,\qquad
\epsilon_n=\bar \epsilon_n \frac{\kappa}{k} =\frac{V_n}{2ka_0}\,,
\eqno(15)
$$
$$
 T_{\delta}= \left ( \begin{array}{cc}
{\cal A} & {\cal B} \\ {\cal B}^* & {\cal A}^* \end{array} \right)\,=
\left ( \begin{array}{cc}
\sqrt{1\!+\!\Delta^2}\, {\rm e}^{i\alpha} & i\Delta
\\ -i\Delta  & \sqrt{1\!+\!\Delta^2}\, {\rm
e}^{-i\alpha} \end{array} \right)\,, \eqno(16)
$$
where the following parameters are introduced
$$
\alpha= -\frac{1}{2}
\left(\frac{k}{\kappa}-\frac{\kappa}{k} \right)\delta=
\frac{\kappa^2-k^2}{2k}\,a_0 \,,
\eqno(17)
$$
$$
\Delta= \frac{1}{2}
\left(\frac{k}{\kappa}+\frac{\kappa}{k} \right)\delta=
\frac{\kappa^2+k^2}{2k}\,a_0 \,.
$$
They are the regular functions of the energy  ${\cal E}=-\kappa^2$
and can be analytically continued to the allowed band, where
$\kappa=i\bar k$ and $\bar k$ is the Fermi momentum in our
system. As usual, we accept that all $V_n$ are
statistically independent, and $\langle
V_n \rangle=0$, $\langle V^2_n \rangle=W^2$. Then the evolution
equations will contain the quantity
$$
\epsilon^2=\langle \epsilon^2_n \rangle=\frac{W^2}{4k^2a_0^2}\,,
\eqno(18)
$$
which is independent of $\kappa$ and trivially continuated
the the allowed band.

\begin{center}
{\bf 3. Evolution equations }
\end{center}

Using the recurrence relation
$$
T^{(n+1)}=T^{(n)}T_\delta T_\epsilon \,,
\eqno(19)
$$
accepting parametrization (1) for  $T^{(n)}$, and
designating parameters of the matrix $T^{(n+1)}$
as $\tilde \rho$, $\tilde \varphi$, $\tilde \theta$,
we have
$$
\sqrt{1\!+\!\tilde\rho}\,{\rm e}^{i\tilde\varphi}
= \sqrt{1\!+\!\rho}\, {\rm e}^{i\varphi}
({\cal A}\!-\!i\epsilon {\cal C}) +\sqrt{\rho}\, {\rm e}^{i\theta}
({\cal B}^*\!+\!i\epsilon {\cal C}^*) \,,
\eqno(20)
$$
$$
\sqrt{\tilde\rho}\,{\rm e}^{i\tilde\theta}
= \sqrt{1\!+\!\rho}\, {\rm e}^{i\varphi}
({\cal B}\!-\!i\epsilon {\cal C}) +\sqrt{\rho}\, {\rm e}^{i\theta}
({\cal A}^*\!+\!i\epsilon {\cal C}^*) \,,
$$
where ${\cal C}={\cal A}\!-\!{\cal B}$. In what follows we
consider the limit
$$
\delta\to 0\,, \quad \epsilon \to 0\,,\quad
\delta/\epsilon^2=const
\eqno(21)
$$
and retain the terms of the first order in $\delta$ and the
second order in $\epsilon$. Squaring the modulus of one of
equations, we have
$$
\tilde\rho= \rho+{\cal D} \sqrt{\rho(1\!+\!\rho)}
+\epsilon^2 (1\!+\!2\rho)\,, \
\eqno(22)
$$
where
$$
{\cal D}= 2(\Delta-\epsilon) \sin{\psi}
-2\epsilon^2  \cos{\psi}\,,
\eqno(23)
$$
and the combined phase variable is introduced
$$
\psi=\theta-\varphi\,.
\eqno(24)
$$
Now let take the product of the second equation (20)
with the complex conjugated first equation
$$
\sqrt{\tilde\rho(1\!+\!\tilde\rho) }\,{\rm e}^{i\tilde\psi}
= \sqrt{\rho(1\!+\!\rho)}\,
\left[\left(1\!-\!2i\alpha\!+\!2i\epsilon\!-\!\epsilon^2 \right)
\,{\rm e}^{i\psi}- \right.
$$
$$  \left.
 -\epsilon^2\,
{\rm e}^{-i\psi}  \right]
+ (1\!+\!2\rho)\,
\left(i\Delta\!-\!i\epsilon\!+\!\epsilon^2 \right) \,.
\eqno(25)
$$
Excluding $\tilde\rho$ using equation (22), we obtain the
relation between  $\tilde\psi$ and $\psi$
$$
\tilde\psi=\psi+2\,(\epsilon\!-\!\alpha)+
R\,(\Delta\!-\!\epsilon)\cos{\psi} +
$$
$$
+R\,\epsilon^2\sin{\psi}+(1\!-\!R^2/2)\,\epsilon^2\sin{2\psi} \,,
\eqno(26)
$$
where
$$
R=\frac{1\!+\!2\rho}{\sqrt{\rho(1\!+\!\rho)}} \,.
\eqno(27)
$$
Using (22), (26) and following the scheme of the papers
\cite{9,15}, we come to the evolution equation for
$P(\rho,\psi)$
$$
\frac{\partial P}{\partial L}=
\left\{\vphantom{\frac{1}{2}}
\!\!-2\Delta\!\sin{\psi} \sqrt{\!\rho(\!1\!+\!\rho\!)} P+
2\epsilon^2\! \sin\!^2{\psi} \rho(\!1\!+\!\rho\!)  P'_{\rho}+
\right.
$$
$$
+\epsilon^2\left[(1\!-\!2\sin^2{\psi})(1\!+\!2\rho)
   -2\cos{\psi}\sqrt{\rho(1\!+\!\rho)} \right] P +
$$
$$ \left.
+2\epsilon^2 \sin{\psi}\left[ \cos{\psi}(1\!+\!2\rho)
   -2\sqrt{\rho(1\!+\!\rho)} \right] P'_{\psi} \right\}'_\rho +
$$
$$
+\left\{(2\alpha\!-\!R\Delta\cos{\psi}) P  \vphantom{\frac{1}{2}}
+\epsilon^2\sin{\psi}(R\!-\!2\cos{\psi}) P
 \right.+
$$
$$ \left.
+\frac{1}{2}\epsilon^2 (2\!-\!R\cos{\psi})^2\, P'_\psi
 \right\}'_\psi \,.
\eqno(28)
$$
The right hand side is a sum of full derivatives, which provides
the conservation of probability.
Integrating over $\psi$, we come to the evolution equation
(4) with parameters \cite{15}
$$
D=2\epsilon^2 \langle\sin^2{\psi}\rangle\,,\quad
\gamma_1 D=\epsilon^2 \langle 1\!-\! 2 \sin^2{\psi}\rangle
\,,\quad
$$
$$
\gamma_2 D=\Delta\langle \sin{\psi} \rangle
-\epsilon^2 \langle \cos{\psi}\rangle\,,
\eqno(29)
$$
which leads to the following result for the "drift
velocity" in Eq.5
$$
v=2 \Delta\langle \sin{\psi} \rangle
+2 \epsilon^2 \langle  1\!-\! \cos{\psi}\rangle
-2 \epsilon^2 \langle\sin^2{\psi}\rangle\,.
\eqno(30)
$$
In the large $L$ limit, the typical values of $\rho$ are
large, and one can set $R=2$. After it the solution of Eq.28
is factorized, $P(\rho,\psi)=P(\rho)P(\psi)$, and the equation
for $P(\psi)$ is splitted off, giving the condition for the
stationary distribution of the phase $\psi$
%
%
$$
\epsilon^2 \left(1\!-\!\cos{\psi}\right)^2 P'_\psi+
\, \epsilon^2 \sin{\psi}
(1\!-\! \cos{\psi})  P+
$$
$$
+ \left(\alpha\! -\! \Delta\cos{\psi} \,
\right) P =C_0 \,,
\eqno(31)
$$
where the constant $C_0$ is fixed by
normalization.\,\footnote{\,Equation (28) is analogous
to Eq.(10.27) in the book \cite{10}, derived in the
framework of the different formalism, so the quantities
entering it are not related clearly with the
transfer matrix parameters.}

\begin{center}
{\bf 4. Transformation properties}
\end{center}

The change of variables in Eq.31
$$
u={\rm tg}\,\psi/2
\eqno(32)
$$
and renormalization of probability  $P\to P(1\!+\!u^2)/2$,
following from $P(\psi)d\psi=P(u)du$, reduce it to the
simple form
$$
u^4 P'_u+(2u^3+a+bu^2) P=C_0 \,,
\eqno(33)
$$
where
$$
a=\frac{\alpha-\Delta}{2\epsilon^2}\,,\qquad
b=\frac{\alpha+\Delta}{2\epsilon^2}\,,
\eqno(34)
$$
or inversely
$$
\alpha=\epsilon^2(b+a)\,,\qquad
\Delta=\epsilon^2(b-a)\,.
\eqno(35)
$$
Equation (33) can be integrated in quadratures, but
this quadrature is practically useless. It is more
effective to investigate the transformation
properties. If $P_{a,b}(u)$ is a solution of Eq.33, then
the following relation is valid
$$
P_{a,b}(u)=s P_{as^3,bs}(su)\,.
\eqno(36)
$$
It can be established, making the change $u=\tilde u/s$ and
reducing the obtained equation to the initial form by
redefinition of parameters $\tilde a =a s^{3}$, $\tilde b
=b s$; then $P_{a,b}(u)$ coincides with
$P_{\tilde a,\tilde b}(\tilde u)$ to the constant factor, which
is established from normalization.
Using the relation
$$
ab=\frac{\alpha^2-\Delta^2}{4\epsilon^4}=
-\frac{\delta^2}{4\epsilon^4}\,,
\eqno(37)
$$
one can see that the scale transformation $a \to a s^3$,
$b \to b s$ leads to renormalization
$\epsilon\to \tilde \epsilon$, where
$$
\tilde \epsilon= \epsilon \, s^{-1}  =
\bar \epsilon \,\frac{\kappa}{k} s^{-1}\,.
\eqno(38)
$$
Substitution of (17) to (34) gives the initial values of the
parameters $a$ and $b$
$$
a=-\frac{\delta}{2\epsilon^2} \frac{k}{\kappa}\,,\qquad
b=\frac{\delta}{2\epsilon^2} \frac{\kappa}{k}  \,,
\eqno(39)
$$
while the relations (35) allow to establish
the change of parameters $\alpha\to \tilde
\alpha$, $\Delta\to \tilde \Delta$  in the result
of the scale transformation
$$
\tilde \alpha=\frac{1}{2}\left( \frac{\kappa}{k} s^{-1} -
 \frac{k}{\kappa} s \right) \delta\,,\quad
\tilde \Delta=\frac{1}{2}\left( \frac{\kappa}{k} s^{-1} +
\frac{k}{\kappa} s \right) \delta \,.
\eqno(40)
$$
Relations (38) and (40) show that transformation of
all parameters $\alpha$, $\Delta$, $\epsilon^2$ entering the
evolution equations reduces to the change
$$
\frac{k}{\kappa} \rightarrow
\frac{k}{\kappa} s \,,
\eqno(41)
$$
which is equivalent to renormalization of the Fermi
momentum in the ideal leads. Inversely, variation of
the properties of the ideal leads results in the scale
transformation of the distribution $P(u)$. However, it is not
sufficient for invariance of parameters $v$ and $D$,
since the simple scaling  $\langle u^n \rangle \to s^{n}\langle u^n
\rangle$ is
a property of only
power averages, while the
actual averages entering to Eqs.\,25,\,26 are not of the
power form
$$
\left\langle \sin^2{\psi} \right\rangle=
\left\langle \frac{4u^2}{(1+u^2)^2} \right\rangle \,,\quad
\left\langle 1\!- \!\cos{\psi} \right\rangle=
\left\langle \frac{2u^2}{1+u^2} \right\rangle \,,
$$
$$
\left\langle \sin{\psi} \right\rangle=
\left\langle \frac{2u}{1+u^2} \right\rangle \,.
\eqno(42)
$$
Meanwhile, invariance of the combination  $\epsilon^2
\langle\sin^2{\psi}\rangle$, determinating the parameter
$D$, demands namely the power scaling for the
first quantity in Eq.42,
$$
\langle \sin^2{\psi} \rangle \rightarrow
s^{2} \langle \sin^2{\psi} \rangle  \,,
\eqno(43)
$$
as it is clear from (38).
This controversial situation is resolved due to specific
properties of the equation (33).

\begin{center}
{\bf 5. Invariance of parameters $D$ and $v$ } \end{center}

Differentiating equation (33), multiplying
it by $u^k$, and integrating in the infinite limits,
we come to the recurrent relation
$$
(k+2)I_{k+3}=b I_{k+2}+ a I_{k}
\eqno(44)
$$
for the integrals
$$
I_{k}=\int u^k P(u) du \,.
\eqno(45)
$$
If the even function $P(u)$ is used in Eq.44, then
a simple relation occurs between the integrals $I_{2k}$,
allowing to express them in terms of  $I_2$,
$$
b I_{2k+2}+ a I_{2k}=0\,\, \rightarrow \,\,
I_{2k+2}=(-a/b)^k I_2\,.
\eqno(46)
$$
If the odd function $P(u)$ is used in Eq.44, then
a simple
relation occurs between the integrals $I_{2k+1}$,
allowing to express them in terms of  $I_1$,
$$
b I_{2k+1}+ a I_{2k-1}=0\,\, \rightarrow \,\,
I_{2k+1}=(-a/b)^k I_1\,.
\eqno(47)
$$
In fact, a solution of equation (33) is not odd, not
even,
and can be represented as a sum
$$
P(u)=P_{odd}(u)+P_{even}(u)\,,
\eqno(48)
$$
where both terms are present inevitably. Using the complete
recurrent relation (44), one can see
$$
I_{4}=(-a/b) I_2 + (4/b) I_5\,,
$$
$$
I_{6}=(-a/b)^2 I_2 - (4a/b^2) I_5 +(6/b) I_7\,,
\eqno(49)
$$
$$
I_{8}=(-a/b)^3 I_2 + (4a^2/b^3) I_5 -(6a/b^2) I_7+(8/b) I_9\,,
$$
etc., so that
$$
I_{2k+2}=(-a/b)^k I_2 +{\rm odd\,\,\, terms}\,,
\eqno(50)
$$
and analogously
$$
I_{2k+1}=(-a/b)^k I_1 + {\rm even\,\,\, terms}\,.
\eqno(51)
$$
Let consider the first average in Eq.42
$$
\left\langle \sin^2{\psi} \right\rangle=
\int \frac{4u^2}{(1+u^2)^2} P(u) du =
$$
$$=
\int \frac{4u^2}{(1+u^2)^2} P_{even}(u) du =
$$
$$
=\int 4u^2 \sum\limits_{k=0}^\infty (-1)^k (k+1) u^{2k}
P_{even}(u) du =
$$
$$=
\sum\limits_{k=0}^\infty 4 (-1)^k (k+1) I_{2k+2} =
$$
$$
= \sum\limits_{k=0}^\infty 4 (-1)^k (k+1) (-a/b)^k I_{2} =
4 I_2 \frac{b^2}{(a-b)^2} \,.
\eqno(52)
$$
Since the averaging function is even, then $P_{odd}(u)$
can be omitted in Eq.48;  expanding the averaging function in
a series, we have the series of integrals $I_{2k}$, containing
the even function $P_{even}(u)$, whose substitution to Eq.50
gives the simple relation (46) between integrals, allowing
to sum the series. Thus
$$
\left\langle \sin^2{\psi} \right\rangle_{a,b}=
 \frac{4 b^2}{(a-b)^2} \int u^2 P_{a,b}(u) du  \,.
\eqno(53)
$$
If the scale transformation (36) is produced in the course of
calculations (52), then
$$
\left\langle \sin^2{\psi} \right\rangle_{a,b}=
\frac{4 \tilde a^2}{(\tilde a-\tilde b)^2} s^{-2 } \int u^2
P_{\tilde a,\tilde b}(u) du  \,,
\eqno(54)
$$
where  $\tilde a= a s^3$, $\tilde b= b s$.
If $s$ is chosen from condition $\tilde a= -\tilde b$,
then the fraction becomes unity, and $s=\kappa/k$, as clear from
(39); then the factor $s^{-2}$  compensates the difference
between $\epsilon^2$  and $\bar \epsilon^2$, and
we come to the result
$$
\epsilon^2 \left\langle \sin^2{\psi} \right\rangle_{a,b}=
\bar\epsilon^2  \int u^2  P_{\tilde a,-\tilde a}(u) du=
\bar\epsilon^2 \left\langle \sin^2{\psi} \right\rangle_{nat} \,.
\eqno(55)
$$
The subscript $nat$ designates the "naturalness" of the
situation with $a=-b$; it corresponds to the condition
$|\kappa|=k$, and according to \cite{15} is
distinguished: in the allowed
band it corresponds to the "natural" ideal leads,
while in the forbidden band it
corresponds to the maximal transparency of interfaces.
In this situation, the quantity  $\epsilon^2$ reduces
to the quantity $\bar\epsilon^2$, determinated by the
internal properties of the system.
The scale transformation $P(u)\to s P(su)$ with large
$s$ results in localization of the distribution in the
small $u$ region, and the averaging function in Eq.52
can be replaced by $u^2$; in this case, the required
invariance is established trivially.

Analogously, for the second average in Eq.42, we obtain
$$
\left\langle 1\!-\!\cos{\psi} \right\rangle_{a,b}=
 \frac{2 \tilde a^2}{\tilde b (\tilde b-\tilde a)^2} s^{-2 }
 \int u^2  P_{\tilde a,\tilde b}(u) du
\eqno(56)
$$
and choosing $s$ from the condition $\tilde a= -\tilde
b$, come to the result
$$
\epsilon^2 \left\langle 1\!-\! \cos{\psi} \right\rangle_{a,b}=
\bar\epsilon^2 \left\langle 1\!-\! \cos{\psi} \right\rangle_{nat}
\,.
\eqno(57)
$$
Now consider the third combination
$\Delta\left\langle \sin{\psi} \right\rangle$ entering
(26). Proceeding in the analogous manner, we have
$$
\left\langle \sin{\psi} \right\rangle=
\int \frac{2u}{1+u^2} P(u) du =
\int \frac{2u}{1+u^2} P_{odd}(u) du =
$$
$$
=\int 2u \sum\limits_{k=0}^\infty (-1)^k
u^{2k} P_{odd}(u) du =
\sum\limits_{k=0}^\infty 2 (-1)^k I_{2k+1} =
$$
$$
=  \sum\limits_{k=0}^\infty 2 (-1)^k  (-a/b)^k I_{1} =
2 I_1 \frac{b}{b-a} \,,
\eqno(58)
$$
so
$$
\left\langle \sin{\psi} \right\rangle_{a,b}=
 \frac{2 b}{b-a} \int u P_{a,b}(u) du  \,.
\eqno(59)
$$
Substituting expressions (17),(39) to (59),
we have
$$
\Delta\left\langle \sin{\psi} \right\rangle_{a,b}=
\delta\cdot \frac{\kappa}{k} \int u P_{a,b}(u) du \,,
\eqno(60)
$$
which gives after the scale transformation (36)
$$
\Delta\left\langle \sin{\psi} \right\rangle_{a,b}=
\delta\cdot \frac{\kappa}{k} \, s^{-1} \int u
P_{\tilde a, \tilde b}(u) du  \,.
\eqno(61)
$$
\begin{figure*}[h]
\centerline{\includegraphics[width=6.5 in]{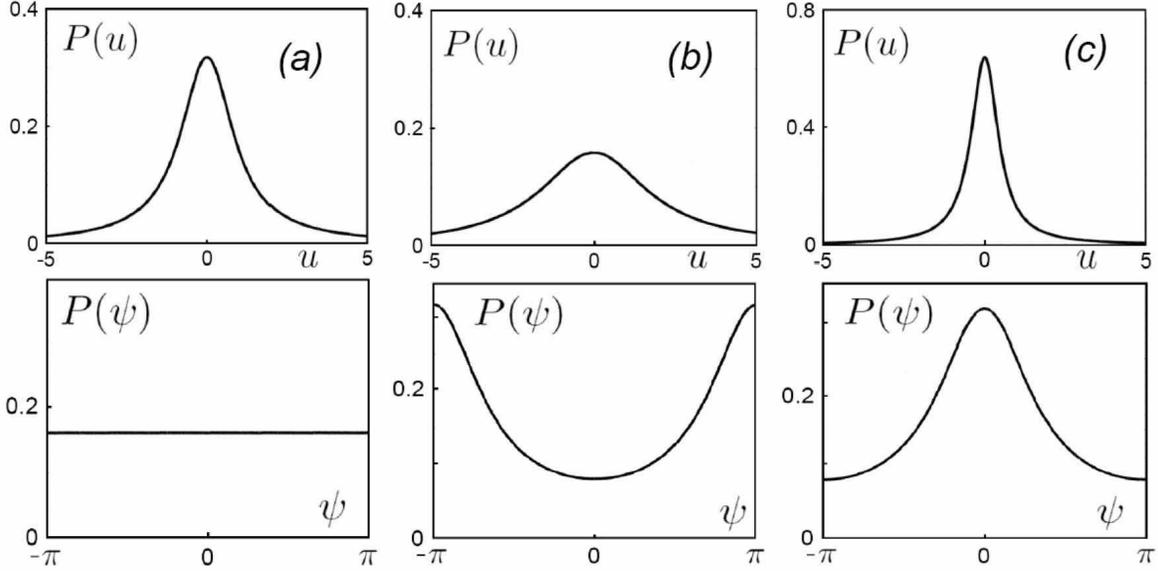}} \caption{
The change of the distribution $P(\psi)$ in the result
of the scale transformation of the function $P(u)$, if the form
of the latter corresponds to the random phase approximation
} \label{fig2}
\end{figure*}
For $s=\kappa/k$ one has equality $\tilde a=-\tilde b$, and
factors $\kappa/k$ and $s^{-1}$ compensate each other, so
$$
\Delta\left\langle \sin{\psi} \right\rangle_{a,b}=
\delta\cdot \left\langle \sin{\psi} \right\rangle_{nat}
\eqno(62)
$$
and we have established invariance of all combinations entering
the expressions (25,26) for $D$ and $v$.

\begin{center}
{\bf 6. Conclusion}
\end{center}

In the present paper we have derived the equation for the
stationary distribution of the phase variable $\psi$,
which determinate the parameters of the limiting
distribution (5) for $P(\rho)$, and establish independence of
these parameters on the boundary conditions, as a consequence
of the transformation properties of the
equation for $P(\psi)$.

In the result of the present analysis we come to a very simple
picture. The phase $\psi$ appears to be a "bad" variable, while
the "correct" variable is $u={\rm tg}\,\psi/2$. The form of the
distribution $P(u)$ is determined by the internal
properties of the system and allows sufficiently strong
variations for the radical change of parameters in the limiting
distribution (5) as a function of the Fermi level (Fig.1).
Variation of properies of the ideal leads, attached to the
system, results in the scale transformation of the function
$P(u)$, which does not affect the values of parameters $v$ and
$D$ due to specific properties of equation (33).

On the qualitative level, variations of the distribution
$P(\psi)$ in the result of the scale transformations of
$P(u)$ are easily predictive and are illustrated in
Fig.2. The distribution $P(\psi)$ is uniform, if $P(u)$ has
a form  $(1/\pi) (1+u^2)^{-1}$ (Fig.2,a), which is valid in
the deep of the allowed band ($\delta\gg\epsilon^2$)
for the "natural" ideal leads ($\kappa^2=-k^2$) and follows
from equation (33) for $a=b$ in the main order in
$\epsilon^2/\delta$.
Widening of the distribution $P(u)$ leads to
localization of $P(\psi)$  near the edges of the interval
$(-\pi,\pi)$ (Fig.2,b), while narrowing leads to
localization of $P(\psi)$ in the middle of the interval
$(-\pi,\pi)$.  (Fig.2,c). However, the rough visual form
of the distribution $P(\psi)$ is not physically substantial
due to invariance of parameters $v$ and $D$
respective the changes shown in Fig.2.

The discussed problems are not restricted by 1D
systems, and analogous difficulties arise in the
studies of the Lyapunov exponents in the
framework of the generalized version \cite{16} of
the Dorokhov--Mello--Pereyra-Kumar equation \cite{17,18}. The
minimal Lyapunov exponent determinates the critical properties of
the Anderson transition (it is clear from the well-known
numerical algorithm, see references in  \cite{16}), and the
analogous hidden symmetry can be essential in the studies of the
latter.


\end{document}